# Stimulated emission does not radiate in a pure dipole pattern

A<small>NDREW</small> E. S. B<small>ARENTINE</small>[1], <small>AND</small> W.E. M<small>OERNER</small>[1,2,*]

[1]*Department of Chemistry, Stanford University, Stanford, California 94305, USA*
[2]*Department of Applied Physics, Stanford University, Stanford, California 94305, USA*
*\*wmoerner@stanford.edu*

**Abstract:** Stimulated Emission (StE) remains relatively unused as an image-forming signal despite having potential advantages over fluorescence in speed, coherence, and ultimately resolution. Several ideas for the radiation pattern and directionality of StE remain prevalent, namely whether a single molecule would radiate StE itself in a pure dipole pattern, or whether its emission direction depends on the driving field. Previous StE imaging has been carried out in transmission, which would collect signal either way. Here, we introduce the StE driving field (the *probe*) at an angle, using total internal reflection to avoid incident probe light and its specular reflections in our detection path. In this non-collinear detection configuration which also collects some fluorescence from the sample, we observe fluorescence depletion even in the spectral window where an increase in detected signal from StE would be expected if StE radiated like a simple classical dipole. Because simultaneous direct measurement of the fluorescence represents a calibration of the potential size of StE were it spatially patterned like a classical dipole emitter, our study clarifies a critical characteristic of StE for optimal microscope design, optical cooling, and applications using small arrays of emitters.

## 1. Introduction

The multitude of refinements of fluorescence microscopy from first attempts at the beginning of the 20[th] century [1] to single-molecule sensitivity in living cells [2-4] has transformed biological exploration over the last hundred years. Meanwhile, stimulated emission (StE), an alternative radiation process theoretically described by Einstein in 1917 [5, 6], has seen relatively little development as an image-forming signal. The sensitivity of fluorescence microscopy can be attributed to the de-coupling of excitation light properties and the resulting spontaneous emission: the dipole radiation pattern of this Stokes-shifted light can be easily separated from its excitation source both spatially and in wavelength, resulting in background-free imaging for good-quality samples. This sensitivity of fluorescence comes at several costs: i) not all emission angles can be collected, ii) the resulting light is spectrally broadband and of random phase, so it is unsuitable for conventional interferometry with a reference source except at low temperatures [7], and iii) the dwell-time in the bleaching-prone excited state is not a simply controlled parameter, but rather a property of the molecule and its environment which sets an ultimate limit on rate of emission.

  The processes of fluorescence and stimulated emission both begin by pumping a molecule in a low electronic state (often the ground state singlet, $S_0$) to an excited state (e.g. $S_1$), as shown in figure 1a. In condensed phase media, after a fast (picosecond) vibrational relaxation, the molecule lingers in the vibrational ground state of this excited electronic state for the excited state lifetime on the order of ns for electric dipole-allowed transitions. The process of StE occurs when the molecule is not allowed to then relax spontaneously by either fluorescence emission or internal conversion, but is instead driven to a lower state (such as a vibrationally excited state of $S_0$) by incident light, radiating light with the same energy as the incident field. For this transition to be favorable, this StE-driving field, the *probe*, is chosen to be red-shifted from the original pumping wavelength to account for the energy lost by vibrational relaxation in the excited and ground states with probability given by Frank-Condon overlap integrals.

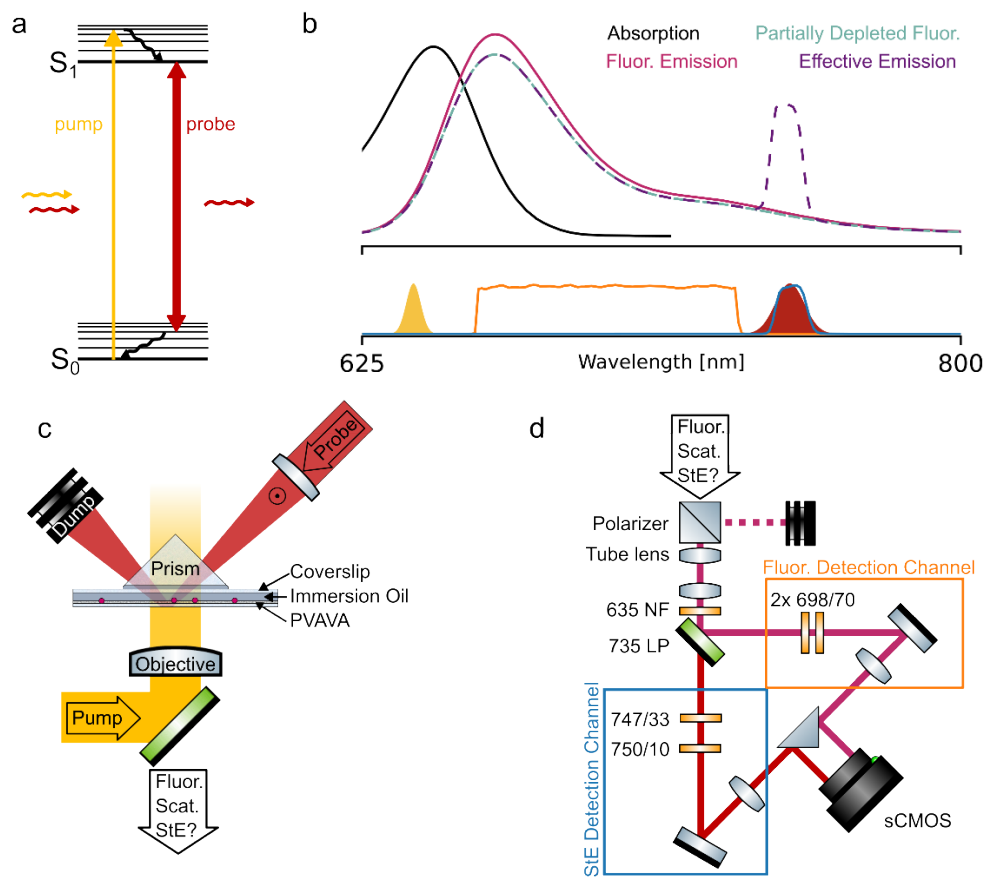

Fig. 1. (a). Jablonski diagram showing the energy levels relevant for stimulated emission (StE), fluorescence emission not shown, and internal conversion is assumed small. Input pump (yellow) and (red, optionally delayed) probe photons are drawn on the side, as is the exciting probe-wavelength StE photon. (b) ATTO 647N absorption (black) and emission spectra (red) for fluorescence (fluor.), partially depleted fluorescence (light blue), and the effective emission spectra when StE is being generated (dashed purple). The pump and probe wavelengths are shown on the lower axis, along with our fluorescence and StE channel transmission spectra (orange and blue, respectively). (c) Diagram of the total-internal reflection (TIR) probe setup. (d) Two-channel detection path diagram. NF: notch filter, LP: long pass

The resonant interaction with the incident light imparts a coherence, implying a fixed phase and wavelength of StE. Rather than emitting fluorescence, with its broad probability distribution of resulting wavelength (see Fig. 1b), the spectral width of StE will be set by the probe. StE competes with fluorescence as a relaxation pathway from the excited state, leading to (partial) depletion of fluorescence across most of the emission spectrum, other than the probe spectral region for which there is an increase in emission signal due to StE gain. This effective emission spectrum is depicted in figure 1b for the highly fluorescent ATTO 647N emitter given a 10% fluorescence depletion effect (simplifying here by ignoring quantum yield and excited-state absorption).

The propagation direction and spatial profile are, however, not necessarily implied by the coherence of StE. The extent to which the structure and orientation of the molecule plays a role has been given little attention, in part because the emission pattern of StE from an extended distribution of emitters, as found in laser gain media, is clearly dominated by the spatial pattern of the probe. Multiple ideas remain prevalent in the literature about the directionality of stimulated emission from a subwavelength emitter, ranging from StE itself radiating in a straightforward dipole pattern [8] to spatially mode-matched directional emission [9, 10], as well as careful theoretical studies where StE arises from the interference between a dipole field and the probe field [11, 12]. Experimental groups targeting stimulated emission imaging have formed images working with different physical pictures for StE propagation, however these experiments have all used a transmitted light configuration, meaning some stimulated emission would reach their detectors regardless of the nature of its propagation.

Here, we present an experiment which generates StE using a driving field which is not incident on our detector (see Fig. 1c), yet features high detection sensitivity for radiation emitted in a dipole pattern. This unique configuration allows us to test whether StE from a sub-wavelength ensemble of emitters can be detected as classical torus-like dipole emission in the far-field or not. We perform our tests using ATTO 647N, a molecule which has been shown to efficiently produce StE, with a small excited state absorption rate at our chosen probe wavelength, such that observed fluorescence depletion serves as a calibration of the amount of StE generated [13]. Imaging sample emission in two spectral windows simultaneously (Fig 1d), we can directly compare the signal modulation due to probe excitation in a fluorescence-only window as well as in a probe-overlapping spectral window. Using ATTO 647N-doped subwavelength polymer beads, we observe only fluorescence depletion in both the fluorescence and StE channels, proving that StE does not radiate into a simple dipole pattern.

## 2. Methods

### 2.1 Probe-TIR microscope

We introduce our pump beam in a typical epifluorescence configuration, from an air-space objective below the sample (see Fig. 1c, Fig. S1, Table S1, and Supplemental Note 1). To stimulate emission from a sample without collecting incident probe light on our detector, we introduce the s-polarized probe at a 45° angle from above and through a prism. After the probe is incident on an index-matched sample layer, it undergoes total internal reflection (TIR) off the bottom coverglass-air interface, and is reflected upward and thereby diverted from our detection path below. The objective below the sample collects dipole radiation: fluorescence emission, probe scatter, and potentially stimulated emission.

A Glan-Laser polarizer is placed in infinity space to attenuate remaining probe scatter collected by the imaging objective, as shown in figure 1d. This scattering signal results from imperfections in the sample itself or in the glass surfaces, and from small refractive index differences between the polymers and the glass. The probe will preferentially drive StE from pumped molecular emitters according to their transition dipole projection along the probe's linear polarization, which is the same orientation nulled by the emission path polarizer. Considering a large ensemble of fixed-orientation dipole emitters, the net field from their combined emission will cancel away from the probe polarization and will tend to be nulled by the polarizer unless symmetry is broken by preferentially pumping dipoles oriented away from the probe polarization. We therefore introduce the pump beam with linear polarization at 45° (see Fig. S2).

The 45° polarization difference between the pump and probe beam reduces the efficiency of generating StE by 1/3. We expect the polarizer to transmit half of the fluorescence emission, which we denote as $T_f = 0.5$, and to transmit a lesser portion of generated StE, $T_s \approx 0.15$. We refine these values to $T_f = 0.57$ and $T_s \approx 0.21$ using measured pump and polarizer orientations (see Fig. S3 and Supplemental Note 2).

A tube lens after the polarizer forms the primary image plane, which is then relayed by a 4f telescope and split into two spectral windows using a dichroic mirror. One spectral window collects fluorescence emission (698/70 nm), the *fluorescence channel*, while the other is matched to the probe beam excitation filtering (750/10 nm), the *StE channel*. Both spectral channels are imaged simultaneously on two regions of the same sCMOS camera.

*2.2 Differential measurement*

To sense an effect due to StE in the presence of background signal from probe scatter, we employ a differential lock-in measurement scheme. We modulate the detected signal by generating two distinct pump-probe relative timings, which either generate stimulated emission (P0 timing) or do not generate stimulated emission (P1 timing), without changing the time-averaged illumination intensity present in either channel. In this way, we can subtract camera frames integrating P1 pulse trains from camera frames integrating P0 pulse trains to remove scatter and fluorescence to obtain a StE difference image. With short pulses (less than the fluorescence lifetime), it is possible to be very efficient at generating StE, using lower intensity than would be required for the same effect size with continuous wave (CW) beams. However, these short pulses serve to synchronize StE events from multiple emitters present in the sample, while long pulses (orders of magnitude longer than the fluorescence lifetime) result in more temporally spread events. We explore both regimes in our experiments.

In the short-pulse regime, we use pulsed lasers with a repetition rate of 76 MHz and pulse widths of 90 and 200 ps for the pump and probe, respectively, as is typical for previous STimulated Emission Depletion (STED) microscopy with these lasers [14]. In the P0 timing (Fig. 2a), the pump creates an excited state population just before the probe beam arrives, generating stimulated emission during the early part of the spontaneous emission excited state lifetime. In the P1 timing (Fig. 2b) the pump arrival time is delayed, so that the probe beam arrives early, when there is no excited state population, and stimulated emission is not driven. Each camera frame integrates over 10 ms of either P0 or P1 pulse trains, and a fast switch toggles between directing P0 or P1 trigger pulses to the pump, synchronously with each camera frame (see Fig. S4 and Table S2), that is at a rate of 100 Hz.

In the long-pulse regime, the pump and probe are both CW lasers, chopped by either an acousto-optic modulator (AOM) or optical chopper, respectively. The camera is synchronized to the probe chopper, and each 56 ms integration time camera frame begins with the arrival of the 28 ms probe pulse. The P0 timing is generated by overlapping a 23 ms pump pulse with the probe pulse, generating stimulated emission (Fig. 2c). The P1 timing is generated by the 23 ms pump pulse arriving after the probe (Fig. 2d). A fast switch toggles on every frame whether the pump AOM driver is given the early P0 or late P1 trigger (see Fig. S5 and Table S2), at a rate of 17.9 Hz.

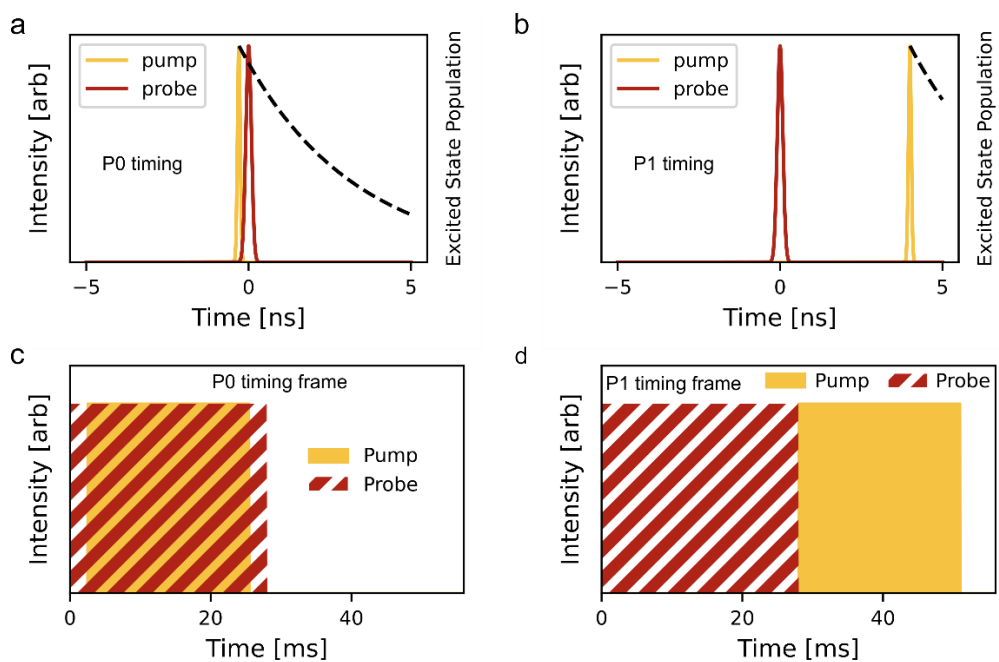

Fig. 2. Pump-Probe pulse timing schemes for differential StE and fluorescence measurements. (a) Pump-probe pulse timing suitable for generating StE (termed, P0), where the pump beam arrives just before the probe beam such that the probe interacts with a large excited-state population. (b) Pump-probe pulse timing P1, where the pump beam arrives after the probe beam, and several 3.5 ns fluorescence lifetimes before the next probe beam pulse arrives (12.5 ns period), such that no stimulated emission is generated. (c) Long-pulse P0 pulse timing, again suitable for generating stimulated emission, using camera frames synchronized to a 50% duty-cycle chopper in the probe beam path. (d) Long-pulse P1 pulse timing, where no stimulated emission is generated but the pump and probe fluences are the same as for a P0 camera frame.

### 2.3 Sample

In order to image a dye known to efficiently radiate StE, and with minimal scatter, we produced covalently labeled ATTO 647N 112 nm diameter poly(methyl-methacrylate) PMMA beads (see Supplemental Note 3). We sonicated and then diluted the beads in 1% Poly(vinyl alcohol – vinyl acetate) (PVAVA) and spin-coated this solution on a #1.5 coverglass which had been plasma-etched with Argon. Using a small drop of immersion oil matched to glass, a second coverglass was then mated on top of the sample. After mounting on the microscope, immersion oil was placed on top of the upper coverslip before lowering the prism into position (see Fig. 1c).

These 112 nm diameter beads are subwavelength in size for our 750 nm probe beam. Given the refractive index of PMMA is 1.48, the maximum emitter spatial separation for these beads is about $0.22\lambda$. This $0.22\lambda$ is an upper bound on the furthest apart two dye molecules can be within the bead and is an over-estimate for the average separation. We regard these beads as a reasonable target to test StE properties in a regime that is not influenced by conventional phasing effects one would expect from a extended distribution of coherently radiating dipoles (which would yield directional radiation [15]).

### 2.3 Series Acquisition and Analysis

Our microscope is controlled by the PYthon Microscopy Environment [16] using acquisition protocols for our short-pulse and long-pulse experiments (see Supplemental Note 4). Our acquisitions begin by stopping live acquisition on the camera and zeroing the divide-by-2 circuit that counts camera frames and toggles P0 and P1 pulse timings for the pump beam. This allows us to label each frame as a P0 or P1 frame throughout the acquisition. The camera is then started, and several camera frames are acquired with the probe shutter open and the pump shutter closed. This is then reversed to acquire several pump-only frames, after which both shutters are opened. Acquisition events are timestamped when the pump and probe shutters are opened or closed, and additional frames are dropped during post-acquisition analysis such that pump-only, probe-only, and both-laser frames do not include frames partially exposed during shutter movement. Image intensities are dark subtracted and converted to units of e⁻/s, ignoring the duty cycle of illumination. The pump-only and probe-only frames are averaged in time. A pixel position corresponding to the center of the bead in each channel is manually selected using the fluorescence spot in the pump-only average image, and used as the center position to crop an 11 x 11 pixel ROI around the bead in each channel.

We calculate the P0 – P1 subtraction first in the time-forward direction, subtracting each P1 frame from the P0 frame that came before it, resulting in a series of frame-pair subtraction images. Leveraging that photobleaching is monotonic in time, we mitigate its effect in our lock-in measurement by additionally calculating the P0 – P1 subtraction in the time reversed direction (subtracting each P1 frame from the P0 frame that came after it) and averaging. The resulting time-balanced P0 – P1 subtraction therefore largely cancels the effect of photobleaching. We use the same bead ROI identified previously to calculate the ROI sum of this P0 – P1 subtraction in time, and additionally average this P0 – P1 subtraction series in time to reconstruct a StE difference image. Finally, we implement a negative control, where we perform the subtraction at half the frequency. This is accomplished by averaging a P0 and P1 frame together, and subtracting the average of the next P0 and P1 frames. This half-frequency negative control P0 – P1 is again performed in the time-forward and time-reversed direction and averaged to mitigate the influence of photobleaching.

*2.4 Expected rate of StE*

Imaging the sample with the same collection optics and direction in both fluorescence-only and probe-overlapping spectral windows, acquired simultaneously on the same camera, provides a useful internal calibration for our data where the common assumption is made that depleted fluorescence photons lead to StE events. For a measured rate of fluorescence depletion, $D_{\text{meas}}$[ e⁻/s], we calculate an expected total rate of StE events in the sample, $S_{\text{abs}}$ [event/s], as

$$S_{\text{abs}} = \frac{D_{\text{meas}}}{\eta \, \Delta\lambda_f \, \text{QE}_f \, C_{\text{obj}} T_f}$$

where $\eta$ is the fluorescence quantum yield, $\Delta\lambda_f$ is the fraction of fluorescence emission detected by our spectral filters, $\text{QE}_f$ is the quantum efficiency of the camera in the fluorescence detection region, and $C_{\text{obj}}$ is the fractional solid angle collected by our objective lens. Note that we ignore the loss from excited state absorption, measured to be small for ATTO647N [13]. We estimate $S_{\text{abs}}/D_{\text{meas}}$ to be 25.8 given $\eta = 0.65$, $\Delta\lambda_f = 0.63$, $\text{QE}_f = 0.83$ [e⁻/event], and $C_{\text{obj}} = 0.2$.

We additionally wish to estimate what the detection rate of StE would be under the hypothesis that it emits into a simple classical dipole pattern, $S_{\text{meas}}^{\text{dipole}}$, which is given by

$$S_{\text{meas}}^{\text{dipole}} = \frac{D_{\text{meas}} \, \Delta\lambda_s \text{QE}_s T_s}{\eta \, \Delta\lambda_f \, \text{QE}_f T_f}$$

where $\Delta\lambda_s$ is the fraction of the probe spectrum transmitted by the StE channel filters, and $\text{QE}_s$ is the camera quantum efficiency in the StE detection window. Note that the objective collection efficiency is shared by both the measured fluorescence depletion and hypothetical dipole StE. The decrease in efficiency of generating StE using a probe 45° misaligned from the pump is already accounted for by calculating a StE rate relative to the measured fluorescence depletion. We estimate $S_{\text{meas}}^{\text{dipole}}/D_{\text{meas}}$ to be 0.59 given $\Delta\lambda_s = 0.75$ and $\text{QE}_s = 0.72$ [e⁻/event].

## 3. Results

We first imaged 112 nm diameter ATTO647N-labeled PMMA beads with the short-pulse train (Fig. 2a, b). Pump-only fluorescence images in both channels are shown in Fig 3a, b, and as expected the signal strength is about 40-fold larger in the fluorescence-only channel than the StE channel (StE ROI sum / Fluor ROI sum: 0.020 measured, 0.025 expected). Probe-only images show some weak probe-excited fluorescence in the fluorescence channel (Fig. 3c) and are dominated by probe scatter in the StE channel (Fig. 3d, note colorbar scales). Looking in time at the bead ROI emission value summed in each channel on each frame (Fig 3e), we see a clear modulation (note the vertical separation between the open and closed circles) which requires the presence of both beams, and calculate a fractional fluorescence depletion modulation of 12.5% (averaged over the first $\tau_b$ (1/e time) of photobleaching) The time-averaged P0 – P1 subtraction image was averaged in time over the first $\tau_b$ (1/e time) of photobleaching. Clear fluorescence depletion can be seen in the fluorescence channel (Fig 3f). Signal of the same (negative) sign, but lesser magnitude is present in the StE channel subtraction image (Fig 3g). The half-frequency P0 – P1 negative control analysis, averaged over the same duration, shows effectively no modulation in either channel (Fig. 3h, i. Colorbars are the same as for Fig 3f, g, respectively).

The fluorescence depletion signal clearly serves as a useful calibration to be able to estimate the expected size of any StE signal, if the StE signal has a dipole emission pattern like fluorescence. Using the ratio of pump-only fluorescence detection between the fluorescence and StE channels, we scale the fluorescence depletion signal (Fig. 3j, orange), to calculate the fluorescence depletion signal we would expect to observe in the StE channel in the absence of any StE gain (Fig. 3j, black). We find excellent agreement between the expected fluorescence depletion in the StE channel, and the measured StE channel subtraction signal (Fig. 3j, blue), with the average difference between the two well within 1 standard deviation of zero ($1.7 \times 10^3$ [e-/s] average difference over the first $\tau_b$, $1.4 \times 10^4$ [e-/s] standard deviation). Given that any hypothetical dipole-patterned StE should result in a large *positive* StE gain signal in that channel (Fig. 3j, purple), we conclude that for the case of a subwavelength-extent ensemble of emitters simultaneously driven to radiate StE, StE is about 139-fold smaller (averaging $\leq 1.7 \times 10^3$ [e-/s] over the first $\tau_b$) than expected for dipole emission (purple curve, $2.3 \times 10^5$ [e-/s] average over the first $\tau_b$), strongly showing that StE does not radiate in a simple dipole pattern.

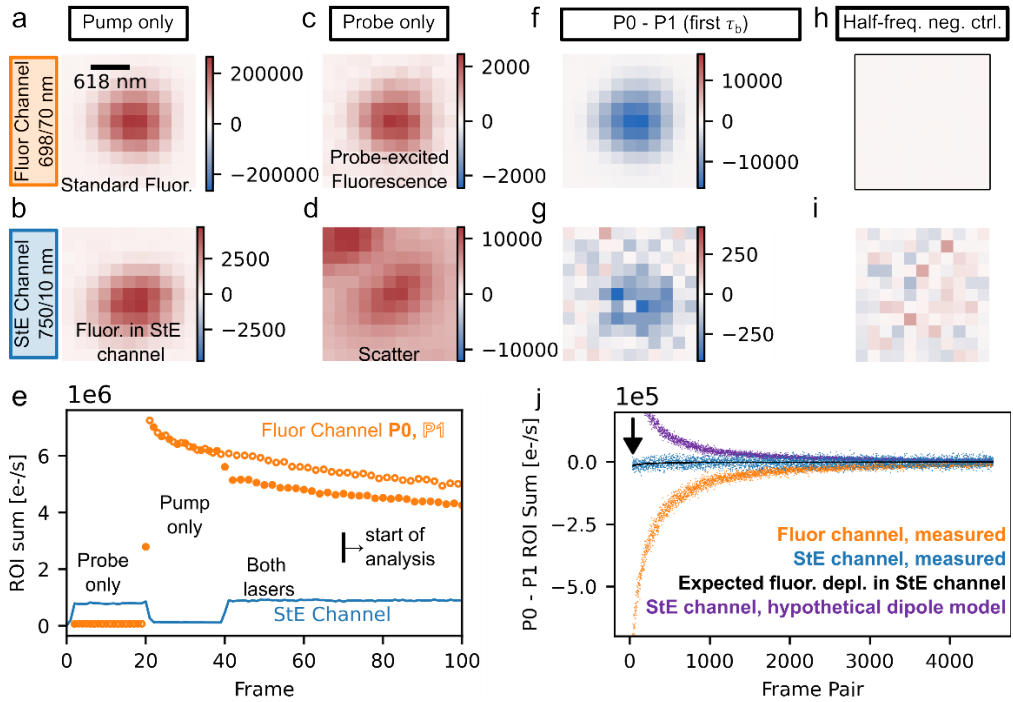

Fig. 3. Typical short-pulse experimental results. A 112 nm ATTO 647N-labeled PMMA bead in the focus of our microscope imaged concomitantly in a fluorescence-only channel (a,c,f,h) and a probe-overlapping spectral channel (b,d,g,i) using ~100 ps pump and probe pulses. Units for each image are in detected photoelectrons/s. (a, b) Pump-only frames show simple fluorescence images of the bead in both channels. (c, d) Probe-only frames detect probe-excited fluorescence in the fluorescence channel (c) and are dominated by probe scatter in the probe-overlapping channel (d). (e) The bead ROI signal in each channel is summed over each of the first 100 frames and plotted. The fluorescence data points are filled or hollow for P0 or P1 frames, respectively. The vertical black line indicates the first time point considered in the following subtraction analysis (f-j). (f, g) $P0 - P1$ subtraction images, averaged over the first $\tau_b$, show fluorescence depletion in both channels. (h, i) Half-frequency negative control subtraction images averaged over the same period show no modulation in either channel. Colorbar scales for panels h and i are the same as f and g, respectively. (h) is outlined in black to show the border despite the low contrast. (j) The $P0 - P1$ subtraction images are summed in each channel over the bead ROI and plotted. An arrow points to the slight negative dip observed in the StE channel data. The expected fluorescence depletion present in the StE channel is plotted (black), as well as the hypothetical StE channel gain for dipole-patterned StE. Similar results occurred for ~10s of other beads and with other samples.

To explore conditions with different collective coherence properties, we repeated our experiment using continuous wave lasers chopped to centisecond pulses (see Fig. 2c,d) and present the results in figure 4. While these long pulses reduce the efficiency with which we can modulate StE (here we achieve 3.6% fractional fluorescence depletion), the emission events are now dramatically spread out in time. The ~5x10$^4$ [e-/s] fluorescence depletion measured at the start of the acquisition before photobleaching (see Fig. 4j), corresponds to an expected 1.3x10$^6$ StE events per second. Under the assumption these events are uniformly random in time, their average separation would be about 0.77 microseconds. With emission events separated an average of about 220 fluorescence lifetimes, we consider this regime to be quasi-single emitter and a valid test of whether StE radiates into a dipole pattern in the absence of ensemble effects. Here, we again see clear fluorescence depletion in the fluorescence channel (Fig. 4c), and this modulation disappears when the subtraction is done at half the frequency

(Fig. 4d), or in the absence of both lasers (Fig. 4i). While the signal-to-noise is degraded relative to the short-pulse case, the subtraction analysis in the StE channel (see Fig. 4g, and blue points in Fig. 4j) clearly does not show strong StE gain. The difference between the StE channel subtraction signal and the expected fluorescence depletion in that channel (Fig. 4j, black) is within 1 standard deviation of zero (190 [e⁻/s] average difference over the first $\tau_b$, 6.3x10$^3$ [e⁻/s] standard deviation), and 104-fold lower than would be expected for a hypothetical dipole emitter (Fig. 4j, purple, 2.0x10$^4$ [e⁻/s] average over the first $\tau_b$). We therefore conclude that even in the absence of ensemble effects, StE does not radiate in a simple dipole pattern.

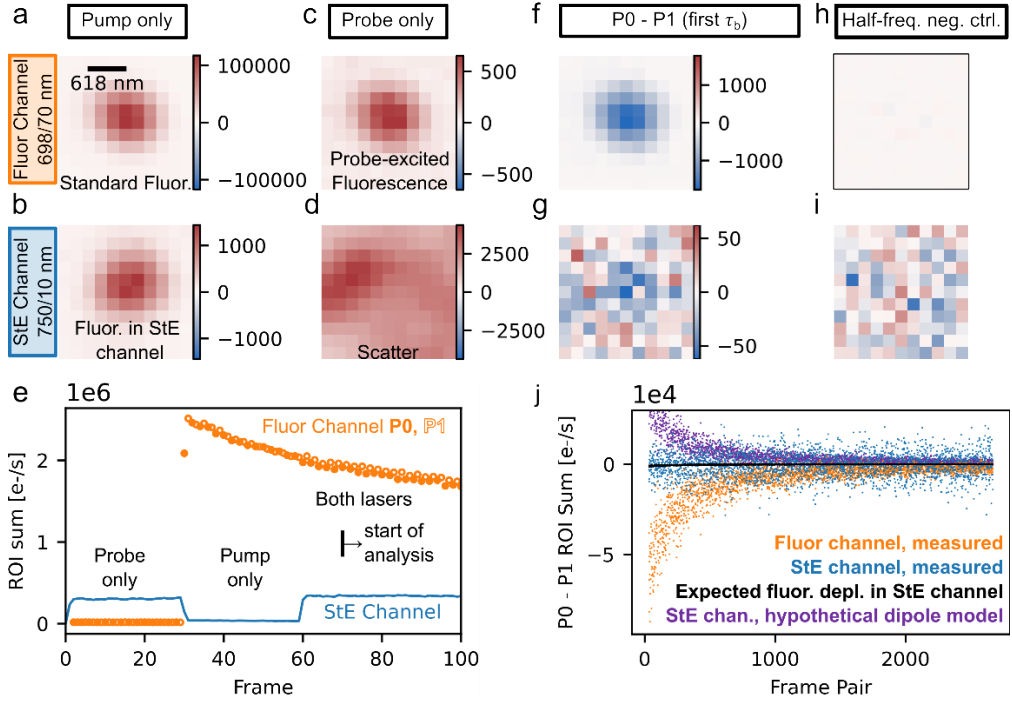

Fig. 4. Typical long-pulse experimental results. A 112 nm ATTO 647N-labeled PMMA bead in the focus of our microscope imaged concomitantly in a fluorescence-only channel (a, c, f, h) and a probe-overlapping spectral channel (b, d, g, i) using ~25 ms pump and probe pulses. Units for each image are in detected photoelectrons/s. (a, b) Pump-only frames show simple fluorescence images of the bead in both channels. (c, d) Probe-only frames detect probe-excited fluorescence in the fluorescence channel (c) and are dominated by probe scatter in the probe-overlapping channel (d). (e) The bead ROI signal in each channel is summed over each of the first 100 frames and plotted. The fluorescence data points are filled or hollow for P0 or P1 frames, respectively. The vertical black line indicates the first time point considered in the following subtraction analysis (f-j). (f, g) P0 − P1 subtraction images, averaged over the first $\tau_b$, show fluorescence depletion in both channels. (h, i) Half-frequency negative control subtraction images averaged over the same period show no modulation in either channel. Colorbar scales for panels h and i are the same as f and g, respectively. (h) is outlined in black to show the border despite the low contrast. (j). The P0 − P1 subtraction images are summed in each channel over the bead ROI and plotted. The expected fluorescence depletion present in the StE channel is plotted in black, as well as the hypothetical StE channel gain for dipole-patterned StE (purple). Similar results occurred for other beads.

## 4. Discussion

Whether StE shares the same spatial profile and propagation direction as the incident field or radiates in a dipole pattern, as spontaneous emission does, has been a question of recent discussion [8, 12]. The coherent properties of StE are appealing as an imaging contrast, however reaching single-molecule sensitivity with StE has so far been impractical using standard fluorophores due to the difficulty in detecting StE appearing at the same wavelength as the probe in the presence of scattering backgrounds. A scheme to image StE without the probe beam being incident on the detector, as it is for standard transmission setups, could greatly benefit achievable sensitivity, but the feasibility of such a scheme would hinge on StE radiating in a manner distinct from the probe.

Treatment of StE from a two-state atom using quantum field theory leads to the result that StE photons share both the wave-vector and polarization of the probe [17]. This is consistent with our experiment results, but is distinct from the classical Lorentz oscillator picture of an oscillating charge on a spring (or in the simple quantum picture, the oscillation of the transition dipole moment) [15, 18, 19]. In these treatments, one would think that the driving field induces a polarization that could radiate (see Supplemental Note 5); however, most authors focus on calculating the energy contained in the interference between the probe and emitted field in the forward direction using the Poynting vector [11, 12, 15]. This interference is contained within the spatial extent of the probe beam, and so could not be detected without shot noise arising from the probe-only component of the Poynting vector.

Here, we configured a measurement, with non-zero probe fields in the sample, to produce stimulated emission fields with an off-axis detection path where the probe field $E_\mathrm{p} \cong 0$. Detecting from such an angle, the only (non-fluorescent) energy flowing towards the detector would be from the dipole-only component of the Poynting vector which could occur from StE or Rayleigh scattering. Since Rayleigh scattering is not detectable in our modulated experiment (see Supplemental Note 5), by calibrating with the observed fluorescence depletion, we find that our results are not consistent with StE radiating in a classical dipole pattern. In other words, StE is not readily detectable in directions away from the probe beam. We hope applications relying on StE from subwavelength emitters benefit from this insight.

**Funding.** National Institutes of Health (R35GM118067).

**Acknowledgments.** The authors appreciate the support and active discussions about this work within the Moerner lab. We additionally would like to thank Adam E. Cohen for sharing thoughts on detecting coherent dipoles and Andrew G. York for engaging discussions about his measurements. This work has been supported in part by the National Institute of General Medical Sciences, USA, United States National Institute of Health, Grant Number R35-GM118067.

**Disclosures.** The authors declare no conflicts of interest.

**Data availability.** Data underlying the results presented in this paper are not publicly available at this time but may be obtained from the authors upon reasonable request.

**Supplemental document.** See Supplement 1 for supporting content.

# Stimulated emission does not radiate in a pure dipole pattern: Supplemental Document

ANDREW E. S. BARENTINE[1], AND W.E. MOERNER[1,2,*]

[1]*Department of Chemistry, Stanford University, Stanford, California 94305, USA*
[2]*Department of Applied Physics, Stanford University, Stanford, California 94305, USA*
*\*wmoerner@stanford.edu*

## Supplemental Figures and Tables

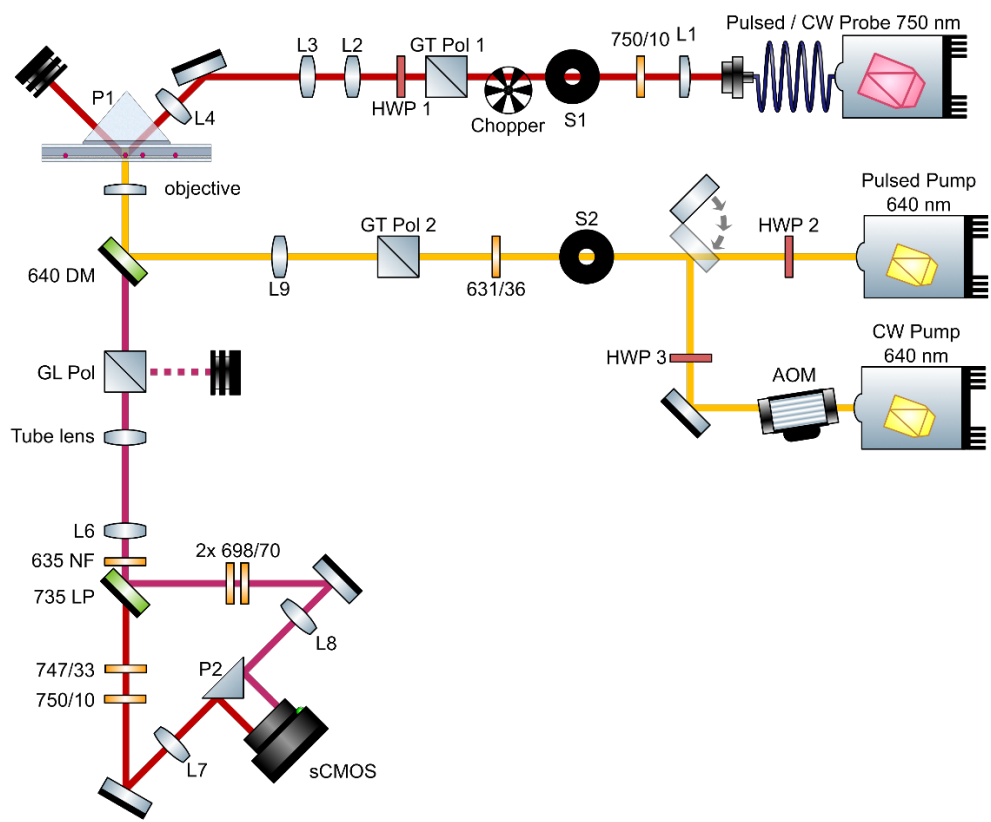

Fig. S1. Optical setup diagram. Abbreviations and component models are listed in Table S1. Silver mirrors were used to preserve polarization control. Not all mirrors are shown in the above diagram. Diagram is not drawn to scale. Wavelength annotations are in nanometers.

| Component | Description | Model; Manufacturer |
|---|---|---|
| **Probe path** | | |
| Probe | Ti:Sapphire, mode-locked (pulsed) or CW, 750 nm | Mira 900 D; Coherent |
| L1 | 10x/0.25 NA, 160 mm tube | M-10X; Newport |
| 750/10 | 750/10 nm bandpass filter | 88-013; Edmund Optics |
| S1 | Shutter | VS14S2T1; Vincent |
| Chopper | Optical chopper | SR540; Stanford Research Systems |
| GT Pol 1 | Glan Thompson polarizer | 10GT04AR.14; Newport |
| HWP 1 | 750 nm zero-order half-wave plate | NHM-100-0750; Meadowlark Optics |
| L2 | f=75 mm, ARC: 650-1050 nm, biconvex singlet | LB1901-B-ML; Thorlabs |
| L3 | f=250 mm | |
| L4 | f=40 mm, ARC: 650-1050 nm, achromatic doublet | AC254-040-B-ML; Thorlabs |
| P1 | N-BK7 Right-Angle Prism, 20mm, -B Coated on legs | PS908L-B; Thorlabs |
| **Pump paths** | | |
| Pulsed Pump | Pulsed 640 nm diode laser (90 ps pulse width) | LDH-IB-640B, PDL-M1-Taiko driver; Picoquant |
| HWP 2 | 647 nm half-wave plate | |
| CW Pump | CW 641 nm diode laser | CUBE 1150205; Coherent |
| AOM | Acousto-optic modulator | 1205-603D; Isomet |
| HWP 3 | 676 nm half-wave plate | Tower Optical |
| S2 | Shutter | VS14S2T1; Vincent |
| 631/36 | 631/36 nm bandpass filter | FF01-631/36-25 |
| GT Pol 2 | Glan Thompson polarizer | B.Halle Nachfl. GmbH |
| L9 | f=502 mm, ARC: 400-700 nm, achromatic doublet | AC254-500-A; Thorlabs |
| **Detection** | | |
| Objective | N PLAN POL 63x/0.8 NA air | 11556056; Leica |
| 640 DM | 640 nm notch dichroic mirror | ZT514/640rpc; Chroma |
| GL Pol | Glan Laser polarizer | from IO-5-730-HP; Thorlabs |
| Tube lens | 200 mm focal length | MXA20696; Nikon |
| L6 | f=200 mm, ARC: 650-1050 nm, achromatic doublet | AC254-200-B-ML; Thorlabs |
| 635 NF | 635 nm laser notch filter | ZET635NF; Chroma |
| 735 LP | 735 nm edge dichroic beamsplitter | FF735-Di02-25x36; Semrock |
| 747/33 | 747/33 nm bandpass filter | FF01-747/33-25; Semrock |
| 750/10 | 750/10 nm bandpass filter | 88-013; Edmund Optics |
| L7 | f=100 mm, ARC: 650-1050 nm, achromatic doublet | AC254-100-B-ML; Thorlabs |
| 698/70 | 698/70 nm bandpass filter | FF01-698/70-25; Semrock |
| L8 | f=100 mm, ARC: 650-1050 nm, achromatic doublet | AC254-100-B-ML; Thorlabs |
| P2 | Right-Angle Prism Mirror, Protected Silver, 25 mm legs | MRA25-P01; Thorlabs |
| sCMOS | Digital CMOS camera | Hamamatsu ORCA-Fusion BT |

Table S1. Optical components shown in Fig. S1.

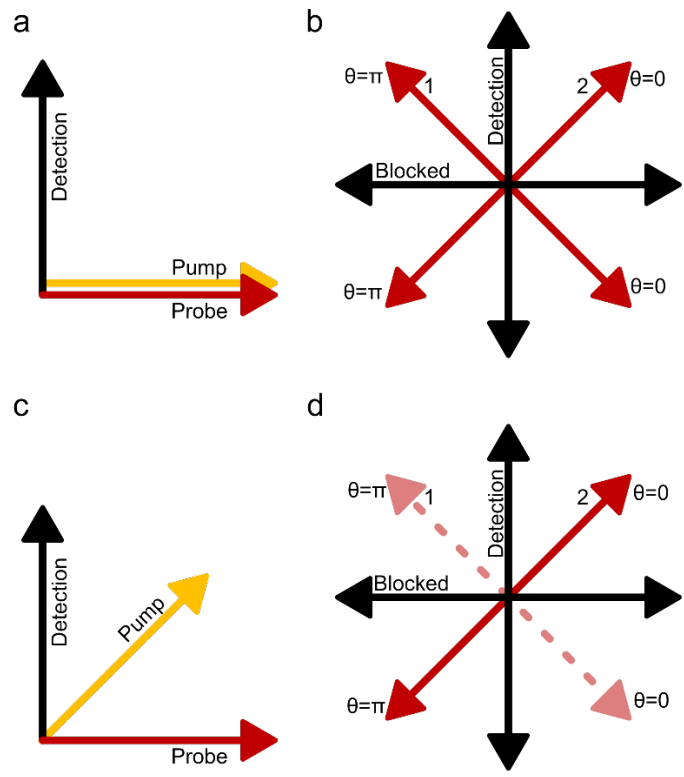

Fig. S2. Excitation and detection polarization schemes. Configuration (i) is depicted in (a), and is not used in our experiments, but is pedagogically important. In configuration (i), the emission path polarizer is set to null the probe polarization, and therefore probe scatter. Matching the pump polarization to the probe as shown would maximize generation of StE. (b) Detection of crossed dipoles for configuration (i). Consider that for a large ensemble of randomly oriented and fixed dipoles, on average every dipole, e.g. dipole 1, has an orthogonal counterpart, dipole 2, as drawn here for a dipole pair oriented 45° away from the probe polarization. If these two dipoles emit StE simultaneously, they radiate coherently and in phase with each other. Their components aligned with the probe, which would add constructively, will be nulled by the emission path polarizer, and their components that would otherwise be detected cancel each other. To avoid this issue (particularly in the short-pulse experiment where we may have simultaneous StE events radiating from the bead), we instead use the configuration (ii) as shown in (c), which takes advantage of the anisotropy in the pumped molecule distribution in the following way. Here, the pump polarization is 45° rotated from the probe polarization. In (d) we consider the same pair of dipoles oriented 45° from the probe polarization, but detected with configuration (ii). Now dipole 1 is orthogonal to the pump polarization, is not pumped to the excited state and therefore does not radiate StE. The broken symmetry in configuration (ii) allows StE to be detected from a large ensemble of coherently emitting dipoles. For an ensemble of randomly oriented and fixed dipoles, it also has the following effects: because the excited state polarization is predominately at 45° from the detection polarization, half of the fluorescence emission is lost at the polarizer. Generating StE (and therefore fluorescence depletion) is 2/3 as efficient as configuration (i), and because excited-state dipoles aligned to the probe preferentially radiate StE, the transmittance of generated StE is about 15% (see Supplemental Note 2 for this calculation).

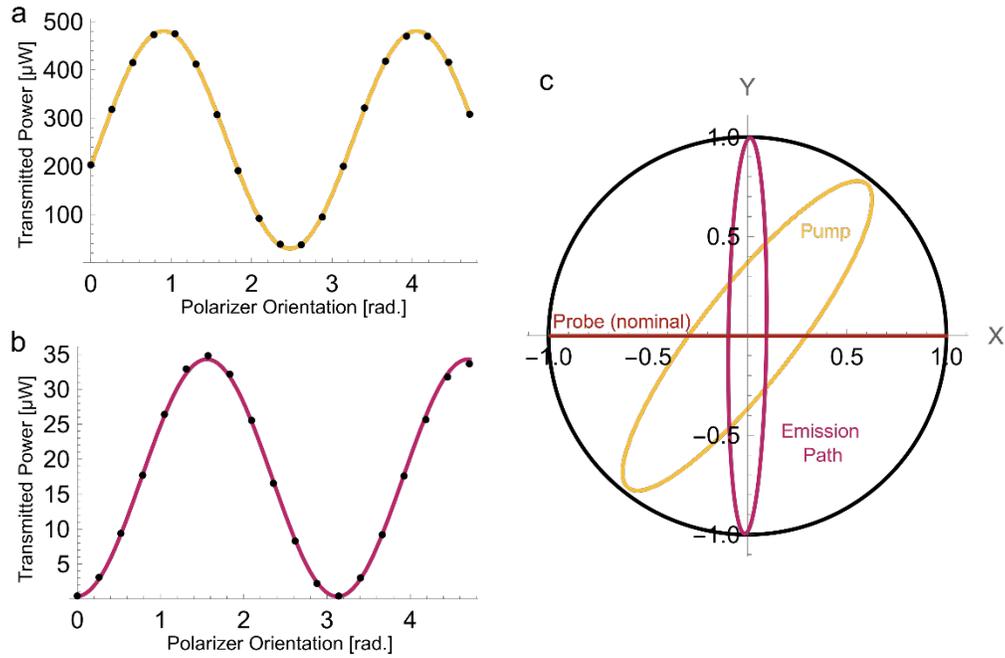

Fig. S3. Sample plane polarization of paths reflecting or transmitting the 640 dichroic mirror. Power transmitted through a thin polarizer in the sample plane was measured for the 640 nm pump beam in its usual epi-illumination configuration, and for the 750 nm probe beam re-routed to pass backwards though the emission path GL polarizer, 640 DM, and objective into the sample plane. The polarizer was then rotated, and the resulting curve fitted to determine the Jones vector representation of the polarization state at the sample plane of the pump (a) and the transmitted emission path light (b). Measuring the light transmitting the system backwards through the GL polarizer into the sample plane is (inversely) equivalent to the transformation linear polarized light in the sample plane would experience while propagating to the GL polarizer. Reflecting or transmitting the 640 dichroic mirror imparts some ellipticity, which is more severe for the pump beam whose linear polarization is at 45° when it reflects off the dichroic. Still, the power extinction ratios of the pump, and transmitted emission path light in the sample plane measured 17 and 79, respectively, and are oriented appropriately with the pump at 52° and the emission path polarizer transmittance at 89°. The power extinction ratio of the thin polarizer rotated during the measurement is 2,133 at 640 nm and 763 at 750 nm. To account for this, the un-rotated Jones matrix used to describe the polarizer is given by $\begin{bmatrix} \sqrt{1-x} & 0 \\ 0 & \sqrt{x} \end{bmatrix}$, where $x$ is the inverse power extinction ratio. We estimate the Jones matrix for the pump in the sample plane to be $\begin{bmatrix} 0.55 - 0.30i \\ 0.78 \end{bmatrix}$, and for 750 nm light transmitting the emission path polarizer into the sample plane to be $\begin{bmatrix} 0.01 - 0.10i \\ 1.0 \end{bmatrix}$. The nominal probe polarization in the sample plane is given by the Jones vector $\begin{bmatrix} 1 \\ 0 \end{bmatrix}$. The Jones vectors are multiplied by a propagator, $e^{i\alpha}$, and $\alpha$ is advanced from 0 to $2\pi$ to display the x- and y-components of the electric field as polarization ellipses in (c), where circular polarization is additionally shown in black as a reference.

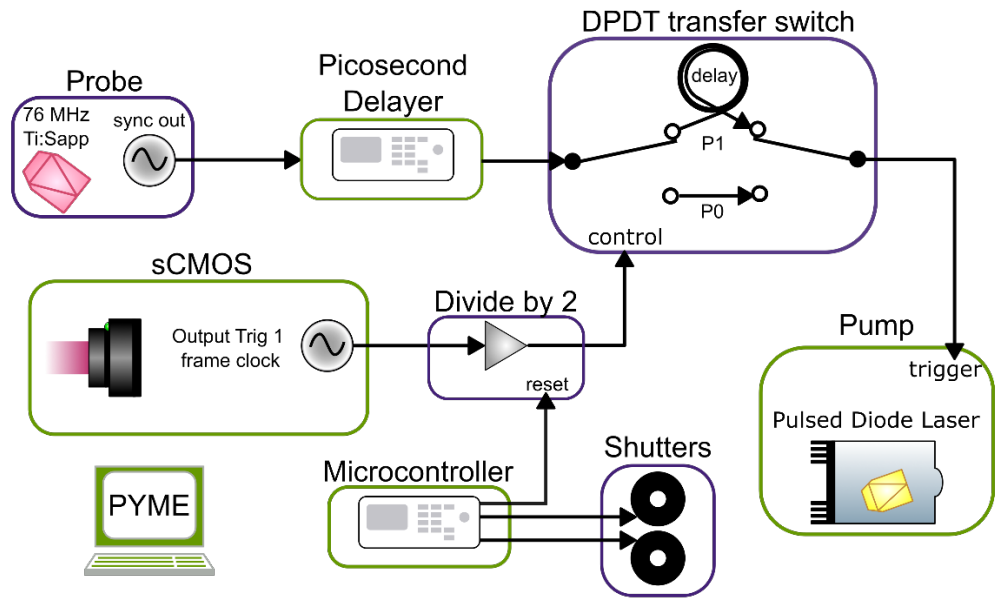

Fig. S4. Short-pulse timing diagram. Our short-pulse experiment has two clocks: the first clock is the Ti:Sapphire probe laser TTL sync-out, which is used to trigger the pump with a controlled delay. This probe train at 76 MHz is adjustably delayed by a solid-state circuit (Picosecond Delayer), before being passed through a double-pole double-throw (DPDT) transfer switch, and into the external trigger input for our pulsed-diode laser pump. Each side of the DPDT transfer switch is bridged by a length of coaxial cable, which is short for one side (generating the P0 delay) and long for the other (generating the P1 delay). The state of the DPDT transfer switch is controlled by the second clock in our system: the sCMOS camera. Each camera frame generates a pulse on the camera output trigger, which is fed through a divide-by-2 circuit. The output of this circuit therefore flops from TTL H to TTL L at the start of each camera frame, in turn changing whether the pump receives a P0 timed trigger from the probe sync-out, or a P1 timed trigger. Green denotes hardware components which communicate with the acquisition computer through PYME.

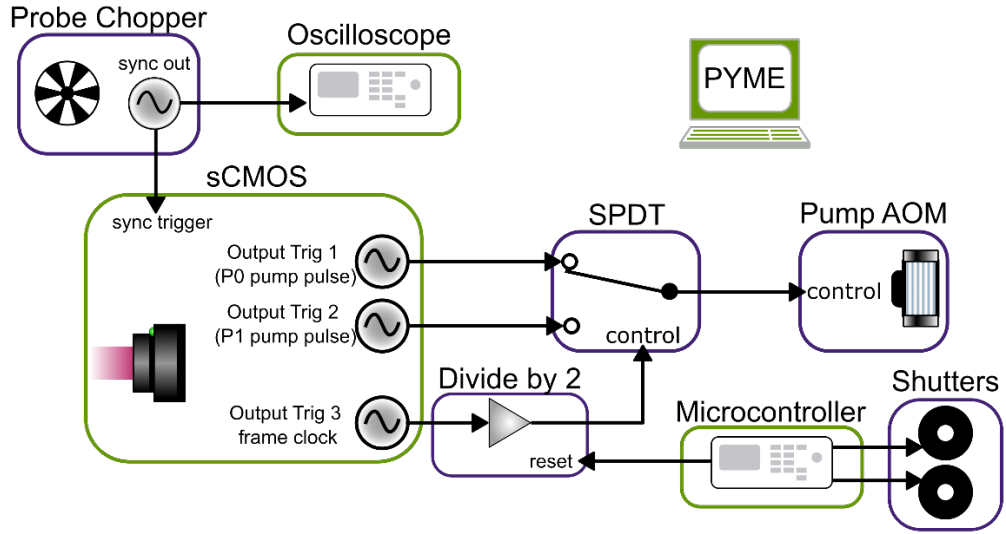

Fig. S5. Long-pulse timing diagram. Our long-pulse experiment has one clock, which is an optical chopper placed in the cw probe beam. The sync-out from this chopper is used as the input trigger for the sCMOS camera, which is running in sync-readout mode. In this mode, the rising edge from each chopper sync-out stops a frame integration and starts the next one, such that there is one probe pulse per camera frame. The camera external trigger delay is set to align the probe beam passing the chopper with the start of each exposure. The camera is then used as a pulse generator. An oscilloscope is used to measure the period of the chopper pulse. The first two output triggers generate a 41% duty cycle pulse, one with only a small delay (which will become the pump P0 pulse), and one with a delay slightly larger than half the chopper period (which will become the pump P1 pulse). Again, an output trigger at the start of each frame is fed into a divide-by-2 circuit. The output of this circuit therefore changes from TTL H to TTL L at the start of each camera frame, and is used to control the state of the single-pole double-throw (SPDT) switch. The P0 and P1 pump pulses generated by the camera are fed into the throws of the SPDT such that only one pulse is fed to the pump AOM control input on a given frame. In this way, each camera frame corresponds to one probe pulse, and either a pump pulse overlapping with it (P0) or not (P1), and the overlap is toggled each frame. Green denotes hardware components which communicate with the acquisition computer through PYME.

| Component | Model; Manufacturer |
|---|---|
| Picosecond Delayer | Micro Photon Devices |
| Divide by 2 | Constructed from DM74LS74; Fairchild Semiconductor |
| SPDT | ZX80-DR230-S+; Mini-circuits |
| DPDT Transfer Switch | 2x ZX80-DR230-S+; Mini-circuits |

Table S2. Additional components shown in timing figures S4 and S5.

**Supplemental Notes**

1. **Instrument**

Our optical instrument is shown schematically in figure S1, and is built on a Nikon TE200 stand. A breadboard is suspended vertically on two damped rods to hold the prism (P1) on top of the sample, and to position a mirror forming the upper half of the periscope which brings the probe above the sample so it can be focused into the prism from above at a 45° angle relative to the sample plane. A silver mirror (Thorlabs P01 coating) is mounted in a filter cube holder routing pump light into and sample emission out of the microscope via the back port.

The Ti:Sapphire laser we use for the probe beam is aligned through several glass rods before coupling into a long single-mode polarization maintaining fiber to stretch the mode-locked pulse length to about 200 ps, as described elsewhere [1]. In addition to the mode-locked operation used in our short-pulse experiments, we use the same laser in our long-pulse experiments by running it CW and using an optical chopper. To run the Ti:Sapphire in CW mode, we open a slit in the cavity slightly and start the laser in CW mode. After collimating the probe beam exiting the other end of the fiber, the beam spectrum and polarization are cleaned using a 750/10 nm bandpass and Glan-Thompson polarizer (GT Pol 1). The probe beam is then expanded using a telescope (lenses L2, L3) before being focused by L4 into the sample through a prism (P1) at a 45° angle relative to the sample plane. In the sample plane, the probe beam is Gaussian in profile, with $\sigma = 1.25$ μm measured perpendicular to the plane of incidence. After irradiating the sample, the probe beam undergoes total-internal reflection (TIR) off the bottom glass-air interface. A zero-order half-wave plate (HWP1) after the polarizer is used to fine-tune the s-polarization of the probe such that scatter is maximally attenuated by the emission path polarizer (GL Pol).

Our pump beam is selected as pulsed or chopped CW for short- and long-pulse experiments, respectively, using a flip mirror. The pump spectrum and polarization are then cleaned by a 631/36 nm bandpass filter and Glan-Thompson polarizer (GT Pol 2), where the latter is rotated 45° relative to both the probe polarization in the sample, and the emission path polarizer (GL Pol). The pump is introduced to the sample through the objective in a standard Kohler illumination epifluorescence configuration using L9 and a 640 nm notch dichroic mirror.

The detection path of the instrument begins with a 0.8 NA air-spaced objective, spec'd for polarization performance. Fluorescence and StE channel light (698/70 nm and 750/10 nm, respectively) both pass the 640 notch dichroic mirror, and transmit the Glan-Laser polarizer (GL Pol) oriented to null scatter in the sample at the probe polarization. The infinity-corrected tube lens is placed outside the microscope body and after the polarizer. A 635 nm notch filter attenuates pump reflections, and a 735 nm dichroic mirror splits the fluorescence and StE channel paths where they are further filtered, each by two bandpass filters. The image plane formed by the tube lens is relayed to the camera by a 4f telescope formed by L6 and L7, and L6 and L8, such that both channels can be brought to the camera with the same nominal magnification (pixel size of 206 nm) using a silver-coated prism (P2).

The probe intensity at the bead during each probe pulse was approximately 70 MW/cm$^2$ for our short-pulse imaging experiment (200 ps probe pulses) shown in Fig. 3, and 580 kW/cm$^2$ for the long-pulse experiment (28 ms probe pulses) shown in Fig. 4. The pump intensity at the bead during each pump pulse was 430 kW/cm$^2$, and 1.3 kW/cm$^2$ for the 90 ps short- and 23 ms long-pulse experiments shown in Fig. 3 and Fig. 4, respectively.

2. **Polarization effects**

In the linear regime where neither pump nor probe transitions saturate, the efficiency of generating StE from an in-plane dipole at orientation $\theta$ is proportional to

$$s = \cos^2(\theta - \theta_{\text{pump}})\cos^2(\theta - \theta_{\text{probe}})$$

where $\theta_{\text{pump}}$ and $\theta_{\text{probe}}$ are the orientations of the pump and probe polarizations, respectively, in the sample plane (recall, the probe is s-polarized). Here, we assume pure linear polarization of the pump and probe. Integrating $\int_0^{2\pi} s \, d\theta$, we find our $\theta_{\text{probe}} = 0$, $\theta_{\text{pump}} = \pi/4$ scheme is 2/3 as efficient at generating StE compared to an experiment with the pump and probe polarizations aligned. The angular distribution of dipoles radiating StE, given by $s$, is symmetrically peaked at $\theta_s = \pi/8$, which would be the polarization angle of their coherent sum. The polarizer, with transmission axis oriented at angle $\theta_{\text{pol}}$, therefore transmits a lesser fraction of StE,

$$T_s = \text{Cos}^2(\theta_s - \theta_{\text{pol}}) \approx 0.15$$

compared to fluorescence, where the latter is determined by the incoherent sum over excited dipole orientations,

$$T_f = \int_0^{2\pi} \text{Cos}^2(\theta - \theta_{\text{pump}}) \text{Cos}^2(\theta - \theta_{\text{pol}}) d\theta = 0.5$$

We can refine these estimates using the measured orientation of the pump and emission path polarizer ($\theta_{\text{pump}} = 52°$ and $\theta_{\text{pol}} = 89°$, respectively. See Fig. S3). This results in $T_f = 0.57$. The distribution of StE-radiating dipoles peaks at $\theta_s = 0.45$ rad. instead of $\pi/8$ rad. (0.39 rad.), and results in $T_s \approx 0.21$. These refined estimates of $T_f$ and $T_s$ are used in calculations such as the expected StE signal magnitude (purple points in Fig. 3j, 4j).

### 3. Sample

We produced ATTO 647N fluorophores covalently attached to 112 nm diameter poly(methyl-methacrylate) (PMMA) beads for our experiments. Amine-labeled PMMA beads (Lab261, PMMA100A) were diluted to 0.1% w/v in 7.4 pH 1x PBS (Gibco, 10010-023) and sonicated for 15 minutes. NHS-ester conjugated ATTO 647N (ATTO-TEC GmbH) was freshly dissolved at 10 mM in anyhydrous dimethylsufoxide (Molecular Probes, D12345) and added to the bead solution for a final dye concentration of at least 0.2 mM, in a working volume of 0.5 mL. The conjugation reaction proceeded for 4 hr at room temperature, with vortex mixing applied several times throughout. The beads were then purified by centrifuging, removing the supernatant, resuspending in 1 mL of 1x PBS, pipette mixing, and then sonicating. This initial purification was repeated at least three times. Finally, tween20 (Fischer Bioreagents, BP337-500) was typically added at 0.1% v/v to help prevent aggregation. UV-vis measurements, blanked on unlabeled beads to account for scatter, suggest an achieved labeling of approximately 7,500 ATTO 647N dyes per bead.

We sonicated and then diluted the ATTO 647N-labeled beads in 1% Poly(vinyl alcohol – vinyl acetate) (PVAVA; Polysciences Inc., #17951, 88% hydrolyzed). To aid in alignment of the probe beam, a separate stock of the PMMA beads were similarly labeled using ATTO 725-NHS, and optionally added to this solution. 60 - 100 uL of this solution was spun onto a 24 x 50 mm #1.5 170 um thick coverglass (Schott D 263 M, acquired from Thorlabs) which had been plasma-etched with Argon. A small drop, 4 uL, of immersion oil (Leica Type-F) was pipetted onto this coverglass before a second plasma-etched coverglass was laid on top of it. This sample was then taped to the microscope stage using Scotch Magic tape. A drop of immersion oil was placed on the top of the coverslip sandwich sample, before lowering a prism attached to a translation stage down onto the sample. This sample configuration is shown schematically in figure 1c.

### 4. Synchronization and Analysis

At the beginning of each series acquisition, the camera is stopped and a list of pre-acquisition tasks in the PYME acquisition protocol is executed serially. This allows us to first zero the camera output triggers, before flashing a voltage onto the reset pin of our divide-by-2 circuit, and then finally enable camera output trigger(s) for hardware-based frame counting during the subsequent acquisition. In the case of the long-pulse experiment, where two camera output triggers generate the pump P0 and P1 pulses, the chopper period is retrieved from an oscilloscope and the output triggers are set with appropriate pulse widths and delays. Pulse timings and our toggle scheme were checked using fast photodetectors and an oscilloscope. The computer time is logged when the camera is started, and the computer time is logged when acquisition protocol tasks are executed so they can be mapped back onto camera frames. In this way, we begin each acquisition by first opening only the probe shutter for several frames, and then switching to only open the pump shutter for several frames, before finally opening both shutters for the remainder of the acquisition. Note that the pulse timing changes synchronized to each frame are occurring throughout each acquisition regardless of whether the shutter is open or closed for a particular beam path.

The signal intensity units of all frames are converted from analog-digital units to e-/s by subtracting the analog-digital offset, multiplying by the gain conversion given by the camera manufacturer, and dividing by the camera integration time. Shutter opening and closing events are mapped into Boolean arrays describing whether frames correspond to pump-only or probe-only illumination. An erosion operation (5 iterations) drops frames from each, and a logical AND operation is used to construct an array describing frames illuminated by both lasers. The pump-only and probe-only frames are retrieved (with an additional 1-frame erosion) and averaged to generate pump-only and probe-only images (cropped bead ROIs shown in Fig. 3a-d and Fig. 4a-d).

Image series of P0 and P1 frames are separately constructed as $P0_i = d_{2i}$ and $P1_i = d_{1+2i}$ where $d_j$ is the $j^{\text{th}}$ camera frame and each $i$ corresponds a frame pair. Half-frequency negative control variants of these image series are constructed as $P0_k^{NC} = 0.5(d_k + d_{k+1})$ and $P1_k^{NC} = 0.5(d_{k+2} + d_{k+3})$ where each $k$ corresponds to a group of 4 camera frames. In this way, each $P0_k^{NC}$ and $P1_k^{NC}$ is an image averaged from an equal number of actual P0 timing frames and P1 timing frames, such that any modulation at our toggle frequency is canceled, and subtracting $P0_k^{NC} - P1_k^{NC}$ should tend to zero.

To create a P0 – P1 difference image, our lock-in measurement, we first create a subtraction series $L_i^{\text{forward}} = P0_i - P1_i$. For a given $i$, $P1_i$ will tend to be more affected by photobleaching than $P0_i$ since it corresponds to a frame which came later in time. This photobleaching effect can be inverted by calculating $L_i^{\text{backward}} = P0_{i+1} - P1_i$ because for a given $i$, $P0_{i+1}$ will be the camera frame just after $P1_i$. We therefore approximately cancel the effect of photobleaching in our lock-in measurement frame-pair subtractions by averaging $L_i^{\text{balanced}} = 0.5(L_i^{\text{forward}} + L_i^{\text{backward}})$. We take the same approach when constructing the time-balanced half-frequency negative control, $H_k^{\text{balanced}} = 0.5(H_k^{\text{forward}} + H_k^{\text{backward}})$, where $H_k^{\text{forward}} = P0_k^{NC} - P1_k^{NC}$ and $H_k^{\text{backward}} = P0_{k+1}^{NC} - P1_k^{NC}$.

The time-balanced P0 – P1 subtraction is plotted in Fig. 3j and Fig. 4j, where the bead ROI in each channel is cropped from $L_i^{\text{balanced}}$ and summed (orange and blue curves for the fluorescence channel and StE channel, respectively). The fluorescence channel ROI-summed frame-pair subtractions are scaled by the measured ratio of fluorescence in each channel from pump-only images at the beginning of the acquisition (Fig 3a, b, and Fig 4a, b) to estimate the expected fluorescence depletion present in the StE channel in the absence of any StE gain (black points in Fig. 3j and Fig. 4j). The fluorescence channel ROI-summed frame-pair subtractions are also scaled by a negative factor of $\frac{S_{\text{meas}}^{\text{dipole}}}{D_{\text{meas}}} = 0.59$ as described in the Methods section 2.4, to estimate what the detected signal of StE gain would be for hypothetical dipole-patterned StE. A negative sign is added because $D_{\text{meas}}$ is fluorescence depletion, or a negative in-phase modulation while the hypothetical detected signal $S_{\text{meas}}^{\text{dipole}}$ would be positive in-phase.

The time-balanced subtraction $L_i^{\text{balanced}}$ does not exist at $i = 1$, or at the last frame of the series, but in any event, we start our lock-in analysis at $i_{\text{start}}$, significantly after the first frame of each acquisition (position marked with a vertical black line in Fig. 3e, Fig. 4e). We also limit our analysis to the first 1/e decay in signal due to photobleaching, with characteristic time $\tau_b$.

We calculate $\tau_b$ by least-squares fitting the fluorescence channel ROI-summed frame-pair subtractions (beginning at $i_{\text{start}}$) to the exponential function $A\left(1 + e^{\frac{-t}{\tau_b}}\right) + c$ where $t$ is time. We find $\tau_b = 6.98$ s and $\tau_b = 32.7$ s for the short-pulse and long-pulse experiments shown in Fig. 3 and Fig. 4, respectively. The subtraction analysis averages in time (Fig. 3f-i, Fig. 4f-i) are taken over the period beginning with $i_{\text{start}}$ and ending $\tau_b$ later.

### 5. Interpreting our results

**Background**

The Lorentz oscillator model of a charged mass on a damped, driven harmonic oscillator can be used in a variety of ways to describe interactions of light with a molecule. While profoundly useful, to allow classical modeling of electric fields radiating from a molecule, the equations of motion must be brought into agreement with quantum mechanical results by e.g. modifying oscillator strengths (see for example, Lasers by Milonni and Eberly, Table 7.1 [2]). We note that it is extremely satisfying to focus on the interference between a Lorentz oscillator dipole scatterer and the probe beam when modeling stimulated emission. This approach, credited to W.E. Lamb and carried out by Cray, Shih, and Milonni in 1982 [3], describes StE as the result from interference of the probe field with the dipole field from the atom, and so naturally recovers directionality in the forward direction expected by QED/QFT. With appropriate oscillator strength, the interference can be found to contain the energy expected for StE [3]. This is clearly a powerful model, however, only the interference term is discussed in most treatments, and the separate dipole-only term is labeled as "Rayleigh scattering or absorption followed by spontaneous emission" [3] and then neglected. This naming of the dipole response as effectively "everything but stimulated emission", despite the oscillator having both an in-phase (with the driving field) dispersive component as well as an out of phase response, can be confusing. With such limited discussion of the dipole-only component in the Poynting vector in the literature, it is understandable that some arrive at a conception of dipole-patterned StE [4, 5]. Therefore, in this paper, we test the limiting case of StE energy predominantly radiating in a dipole pattern.

**Modulated Signal**

In a probe-avoiding detection scheme, under the hypothetical dipole model of StE, the measured intensity is expected to be:

$$I_{\text{side}} \propto |E_R|^2 + |E_{\text{StE}}|^2 + |E_F|^2 + 2(E_R \cdot E_{\text{StE}} + E_R^* \cdot E_{\text{StE}}^*)$$

where we consider the dipole-patterned fields $E_{\text{StE}}$, arising from StE, $E_F$, arising from fluorescence emission, and $E_R$, arising from Rayleigh scatter.

In our experiments, we modulate the relative timing of pump and probe pulse arrivals such that camera frames integrate over pulse trains where the probe is incident on a pumped population of molecules which may be in the excited state (P0 timing) or an un-pumped population of molecules in the ground electronic state (P1 timing). Rayleigh scattering, fluorescence, and StE signals are all affected by this modulation because: i) Rayleigh scattering off an excited-state molecule has a slightly different cross-section than scattering off a ground state molecule, ii) fluorescence is a competing relaxation mechanism from the excited state and is therefore depleted by generation of StE, and iii) StE is not produced by the P1 pulse train. Subtracting $I_{\text{side}}^{P0} - I_{\text{side}}^{P1}$ therefore yields:

$$I_{\text{side}}^{P0} - I_{\text{side}}^{P1} \propto \left|E_R^{P0}\right|^2 - \left|E_R^{P1}\right|^2 + |E_{\text{StE}}|^2 + \left|E_F^{P0}\right|^2 - \left|E_F^{P1}\right|^2 + 2(E_R^{P0} \cdot E_{\text{StE}} + E_R^{*\,P0} \cdot E_{\text{StE}}^*) \quad (1)$$

For typical fluorophores, the Rayleigh scatter cross-section is orders of magnitude smaller than the stimulated emission cross-section in the excited state. Considering $E_R \ll E_{\text{StE}}$, one finds that

$$I_{\text{side}}^{\text{P0}} - I_{\text{side}}^{\text{P1}} \propto |E_{\text{StE}}|^2 + |E_F^{\text{P0}}|^2 - |E_F^{\text{P1}}|^2$$

The negative signal of fluorescence depletion, $|E_F^{\text{P0}}|^2 - |E_F^{\text{P1}}|^2$, is a measure of excited state depopulation due to generation of StE. In our experiments, we find that $I_{\text{side}}^{\text{P0}} - I_{\text{side}}^{\text{P1}} \approx |E_F^{\text{P0}}|^2 - |E_F^{\text{P1}}|^2$, placing bounds on the potential magnitude of $|E_{\text{StE}}|^2$ which are not compatible with a dipole emission pattern for StE. Essentially, we have shown that StE is instead directional, depending on the probe.

Looking at equation 1 and taking $E_{\text{StE}} = 0$, it is apparent that our measurements also bound the magnitude of excited-state Rayleigh scattering, $|E_R^{\text{P0}}|^2 - |E_R^{\text{P1}}|^2$, because then
$$I_{\text{side}}^{\text{P0}} - I_{\text{side}}^{\text{P1}} \propto |E_R^{\text{P0}}|^2 - |E_R^{\text{P1}}|^2 + |E_F^{\text{P0}}|^2 - |E_F^{\text{P1}}|^2$$
Not observing deviation from fluoresce depletion in the probe-overlapping spectral channel indicates that $|E_R^{\text{P0}}|^2 - |E_R^{\text{P1}}|^2 \ll |E_F^{\text{P0}}|^2 - |E_F^{\text{P1}}|^2$, i.e. that fluorescence depletion is much larger than excited-state Rayleigh scatter for our experimental parameters.

## Ensemble effects

For $N$ coherently radiating dipoles, sub-wavelength in extent, the resulting electric field magnitude scales with $N$, and therefore the intensity scales with $N^2$ [6]. This effect is not limited to elastic scattering processes, and is a notable feature of superradiant fluorescence emission possible in conditions where an excited ensemble of emitters does not experience inhomogeneous spectral broadening [7]. While this $N^2$ scaling is not expected for the fluorescence signal in our room-temperature measurements, it would be expected for Rayleigh scattering, and would have additionally been expected for StE under the (invalidated) hypothetical dipole model. Given that this coherent signal enhancement could rescue the otherwise weak effect of excited-state Rayleigh scattering [8] which we did not readily observe, it is interesting to qualify our observation by estimating the size of the excited-state ensemble in our experiments. We estimate the rate of pumping to the excited state, $X_{\text{abs}}$, using the measured fluorescence emission rate, $F_{\text{meas}}$ during the P1 timing camera frames where fluorescence is not quenched by stimulated emission as:

$$X_{\text{abs}} = \frac{F_{\text{meas}}}{\eta \, \Delta\lambda_f \, \text{QE}_f \, C_{\text{obj}} \, T_f}$$

Our short-pulse experiment begins with $F_{\text{meas}} \approx 6 \times 10^6 [e^-/s]$, and therefore $X_{\text{abs}} \approx 1.5 \times 10^8$ [events/s]. Given the laser repetition rate of 76 MHz, this amounts to an average of about 2.0 molecules pumped per pulse, indicating an experimental regime where excited-state Rayleigh scatter is unlikely to be dramatically enhanced by ensemble effects.